\documentclass[superscriptaddress,reprint,longbibliography]{revtex4-1}
\usepackage{amssymb,bbold,dsfont}
\usepackage{amsmath}
\usepackage{color}
\usepackage{graphicx}
\usepackage{marvosym}
\usepackage{ulem}
\usepackage{notoccite}
\usepackage{comment}
\usepackage{pifont}
\usepackage{bbm}
\usepackage{graphicx,tabularx}
\usepackage{dcolumn}
\usepackage{bm}
\usepackage{xcolor}
\usepackage{marvosym}
\usepackage[breaklinks=true, colorlinks,citecolor=blue,linkcolor=blue,urlcolor=blue]{hyperref}
\usepackage{titlesec}

\usepackage[english]{babel}
\def\be{\begin{equation}}
	\def\ee{\end{equation}}
\def \bea{\begin{eqnarray}}
	\def \eea{\end{eqnarray}}
\def \nn{\nonumber}

\begin{document}

	\title{Band geometry induced electro-optic effect and polarization rotation}
	\author{M. Maneesh Kumar}
	\thanks{M.M.K. and S.S. contributed equally to this manuscript.} 
	\affiliation{Department of Physics, Indian Institute of Technology Kanpur, Kanpur 208016}
	\author{Sanjay Sarkar}
	\thanks{M.M.K. and S.S. contributed equally to this manuscript.} 
	\affiliation{Department of Physics, Indian Institute of Technology Kanpur, Kanpur 208016}
	\author{Amit Agarwal}
	\email{amitag@iitk.ac.in}
	\affiliation{Department of Physics, Indian Institute of Technology Kanpur, Kanpur 208016}

\begin{abstract}
Electric field-induced modulation of the optical properties is crucial for amplitude and phase modulators used in photonic devices. Here, we present a comprehensive study of the band geometry-induced electro-optic effect, specifically focusing on the Fermi surface and disorder-induced contributions. These contributions are crucial for metallic and semimetallic systems and for optical frequencies comparable to or smaller than the scattering rates. We highlight the importance of the quantum metric and metric connection in generating the electro-optic effect in parity-time reversal ($\mathcal{PT}$) symmetric systems such as CuMnAs. Our findings establish the electro-optic effect as a novel tool to probe band geometric effects and open new avenues to design electrically controlled efficient amplitude and phase modulators for photonic applications. 

\end{abstract}

\maketitle
\section{Introduction}

Electro-optical effects showcase the remarkable ability of quantum materials to transform their optical properties under the influence of an applied dc electric field. Specifically, for linear electro-optical (EO) or Pockels effect \cite{Ching1982, boyd2003, Chen2014, Li2020, Wang2018}, the change in the refractive index (RI) is proportional to the strength of the applied dc electric field. This phenomenon underpins the functionality of EO materials as amplitude and phase modulators, which enable electrical tunability and manipulation of the optical parameters such as amplitude, phase, and polarization \cite{goldstein1986, Ernesta1999, Cui2017}, for a variety of applications in photonic devices. 
There is immense interest in the exciting phenomena of EO effects and polarization rotation in optical experiments \cite{Gibson2018, Liu2003, Luo_2020, Ye2023}. Earlier studies have focused on material systems such as organic materials \cite{Chelmus2021, Sun2021}, liquid crystals \cite{Blinov1975}, chiral topological semimetals \cite{Ma2015, Sonowal2019, Singh2018}, zinc blends \cite{namba1961}, certain trigonal and piezoelectric crystals \cite{Cheng2019, Fu2000}, Ferroelectric oxide materials \cite{Jiang2020}, Si$_3$N$_4$ platforms \cite{Alexander2018}, silicone waveguides \cite{Berciano2018}, and on certain stoichiometric samples \cite{Fujiwara1999,peterson1964, Kaminow1967, Wang2022}. The EO modulators based on epitaxial BaTiO$_3$ integrated on silicon \cite{Li2019} play a pivotal role in optical communication. 

Despite extensive material-based studies of EO effects, a comprehensive theoretical exploration of the phenomena of EO effect induced by nonlinear optical susceptibilities is still lacking. Specifically, the contributions to the nonlinear optical susceptibilities induced by Fermi surface effects and scattering with impurities have not been explored earlier. These contributions are essential for metallic and semimetallic systems and for exploring low-frequency EO regimes.   Recently, this effect has been studied in chiral topological semimetals \cite{Li2022}, revealing that the second-order response includes contributions from the shift current, injection current, and nonlinear anomalous current induced by the Berry curvature. Here, we develop a generalized theory of the linear EO effect, which works for all systems without any symmetry constraints. Using the density-matrix based quantum kinetic theory framework, we calculate the second-order optical response at finite optical frequencies. We retain all Fermi surface and Fermi sea contributions and disorder effects to identify six unique contributions to the linear EO effect. We denote these contributions as shift, injection, Drude, anomalous, higher-order pole, and double resonant contributions~\cite{Bhalla2022}. As an example to demonstrate our work, we focus on an antiferromagnetic material CuMnAs, which breaks parity ($\cal P$) and time-reversal symmetry ($\cal T$) but is ${\cal P}{\cal T}$ symmetric. 
In a $\mathcal{PT}$ symmetric system, the Berry curvature and the symplectic connection--the imaginary parts of the quantum geometric tensor (QGT), and the quantum geometric connection (QGC) vanish. 
We show that the dominant contribution in the linear EO effect in $\mathcal{PT}$ symmetric systems emerges from the quantum metric and the metric connection, the real parts of QGT and QGC, respectively. More interestingly, we show that the dominant EO effect arises for optical energies lesser than the chemical potential. This highlights that the Fermi surface effects play a significant role in EO effects.

Our paper is structured as follows: In Sec. \ref{section2}, we 
develop the framework for understanding the EO effect starting from Maxwell's equations. In Sec. \ref{section2.5}, we explore the linear EO effect, focusing on the relationship between birefringence and the applied static electric field.  In Sec. \ref{section3}, we introduce the distinct band geometric contributions to optical susceptibilities, capturing the Fermi sea, Fermi surface, and disorder effects. We demonstrate our results in Sec.~\ref{section4}, for a model system of  CuMnAs, a $\mathcal{PT}$ - symmetric material, with broken $\mathcal{P}$ and $\mathcal{T}$- symmetry. Finally, in Sec.~\ref{endgame}, we summarise our work.

\section{The electro-optic effect}\label{section2}

In this section, we employ Maxwell's equations to calculate the modulation of frequency-dependent refractive indices in a medium subjected to a dc electric field. The spatio-temporal evolution of the electric field propagating in a medium is given by \cite{griffiths2017, jackson2012},
\begin{equation}{\label{Maxwell eq for E vector}}
{ \nabla}^2 {\bm E (\textbf{r},t)} = \frac{\partial}{\partial t} \left[ \mu {\bm j}_f + \epsilon \mu \frac{\partial {\bm E (\textbf{r},t)}}{\partial t} \right]~.
\end{equation}
Here, $\epsilon$ ($\mu$) is the static permittivity (permeability) of the medium, and ${\bf j}_f$ is the free charge current density. We consider a light beam propagating with wave vector $\bf{q}$ along the $z$-axis ($\hat{z}$) with its polarization confined to the $x$-$y$ plane. An external bias voltage (static or slowly varying) is applied along the $y$-axis, resulting in a dc electric field $\textbf{E}^d = E_y^d ~ \hat{y}$. 
Thus, for a monochromatic light of frequency $\omega$, the dominant electron current oscillating with a frequency $\omega$ and linear in the optical field strength $E(\omega$), is given by \cite{boyd2003, ashcroft1976}, 
\bea 
j_a(\omega)=\sigma_{ab}(\omega) E_b(\omega) + \sigma^{(2)}_{abc}(\omega; \omega, 0) E_b(\omega) E^d_c(0)~.\label{nl_j}
\eea 
Here, the first term denotes the linear conductivity, and the second term represents the NL current contribution responsible for the linear electro-optic effect. In Eq.~\eqref{nl_j}, the summation over repeated indices is implied. 
We will focus on the propagating plane wave solutions of Eq.~\eqref{Maxwell eq for E vector}, which have the form~$\textbf{E} (\omega) = \textbf{E}_0 e^{i(\textbf{q}\cdot \textbf{r} - \omega t)}$. Using $\textbf{E} (\omega)$, and Eq.~\eqref{nl_j} in Eq.~\eqref{Maxwell eq for E vector}, we obtain,
\begin{equation}
    q^2 E_a = \frac{ \omega^2}{c^2} \left[ \frac{i \sigma_{ab} E_b}{\omega \epsilon_0} + \frac{i\sigma^{(2)}_{abc}E_bE_c}{\omega \epsilon_0} +  E_a \right]~.
\end{equation}
Here, we have used $E_{a(b)} = E_{a(b)} (\omega)$ for brevity. Defining the finite frequency dielectric tensor \cite{cohen2016}, $\epsilon_{aa}(\omega) = 1+ \frac{i}{\omega\epsilon_0}\sigma_{aa}(\omega)$, we can express the above equation as, 

\begin{equation}\label{Eqn:complex_RI}
     \left[\left(\frac{cq}{\omega}\right)^2 - \epsilon_{aa}(\omega)\right] E_a = \frac{i}{\omega \epsilon_0} [\sigma_{ab} E_b (1 - \delta_{ab}) 
   + \sigma^{(2)}_{abc} E_bE_c]~. 
\end{equation}
We define the complex RI, $\tilde{N} = (cq/\omega)$ and $N^2_{aa} = \epsilon_{aa}(\omega)$. The corresponding frequency-dependendent susceptibility tensors are given by $\chi^{(2)}_{abc}(\omega; \omega,0) = \frac{i}{\omega \epsilon_0}\sigma^{(2)}_{abc} (\omega; \omega,0)$ and $\chi_{ab}(\omega) = \frac{i}{\omega \epsilon_0}\sigma_{ab} (\omega)$. Expressing Eq.~\eqref{Eqn:complex_RI} in terms of these, we obtain, 
  \begin{equation} \label{max eqn in N}
     (\tilde{N}^2 - N^2_{aa}) E_a = \chi_{ab}(\omega)E_b(1-\delta_{ab}) + \chi^{(2)}_{abc}(\omega; \omega,0) E_b E_c~.
  \end{equation}
Explicitly writing Eq.~\eqref{max eqn in N} for $E_x$ and $E_y$ in a matrix form, we have, 
   \begin{equation} \label{N_matrix_eqn}
    \begin{pmatrix}
    \tilde{N}^2-N_{xx}^2 - \chi^{(2)}_{xxy}E^d_y && -\chi_{xy} - \chi^{(2)}_{xyy}E^d_y \\
    -\chi_{yx} - \chi^{(2) ~ }_{yxy}E^d_y && \tilde{N}^2 - N_{yy}^2 - \chi^{(2)}_{yyy}E^d_y 
     \end{pmatrix}
     \begin{pmatrix}
     E_x \\
     E_y
     \end{pmatrix} = 0 ~.
\end{equation}
For simplicity of notation and without loss of generality, we express $N_{xx} = N$ and $N_{yy} = N \sqrt{1+ \eta^2}$, where $\eta$ 
is a parameterizing factor. 

To have a non-trivial solution for $E_x~ {\rm and}~ E_y$, the determinant of the matrix must vanish. This yields the following condition,  

\begin{widetext}
\begin{equation}
        (\tilde{N}^2-N^2)^2 - (\tilde{N}^2-N^2)[\eta^2 N^2 + (\chi^{(2)}_{yyy} + \chi^{(2)}_{xxy})E^d_y] + \chi^{(2)}_{xxy}E^d_y(\eta^2 N^2 + \chi_{yyy}E^d_y) - (\chi_{xy} + \chi^{(2)}_{xyy}E^d_y)(\chi_{yx} + \chi^{(2)}_{yxy} E^d_y)=0 ~.
\end{equation}
\end{widetext}
Solving the equation for roots of $(\tilde{N}^2-N^2)$, we obtain   the complex RI $\tilde {N}$ to be  
\begin{equation} \label{Ntilde_solved exprn}
    \tilde{N}_{\pm} = \sqrt{N^2 + \Theta_{\pm}}, ~ \text{where} \hspace{1mm}   \Theta_{\pm} \equiv  \frac{1}{2}[\zeta \pm \sqrt{\zeta^2 - 4 \xi}] ~,
\end{equation}
is a dc electric field and frequency-dependent function. Here, we have defined, 
\begin{eqnarray}\label{eq_zeta}
    \zeta &=& [\eta^2 N^2 + (\chi^{(2)}_{yyy} + \chi^{(2)}_{xxy})E^d_y]~,
\end{eqnarray}
and,
\begin{eqnarray}\label{eq_xi}
    \xi &=& \chi^{(2)}_{xxy}E^d_y (\eta^2 N^2 + \chi^{(2)}_{yyy}E^d_y) \nonumber \\
    &-& (\chi_{xy} + \chi^{(2)}_{xyy}E^d_y)(\chi_{yx} + \chi^{(2)}_{yxy} E^d_y)~.
\end{eqnarray}
Note that for $E^d_y \to 0$, we have $\zeta \to \eta^2 N^2$, and $\xi \to |\chi_{xy}|^2$. The complex RI can be tuned and controlled by varying $E^{d}_y$. 
$\tilde {N}_{\pm}$ corresponds to two solutions for the effective propagating wavevectors, $\bm{q}_{\pm} = (\omega \tilde{N}_{\pm}/c) ~ \hat{q}_\pm,$ of the optical field. The complex RI can be expressed in terms of the refractivities ($n_{\pm}$) and the extinction (or absorption) coefficients ($\kappa_{\pm}$), $N_{\pm} = n_{\pm } + i \kappa_{\pm}$ \cite{visnovsky2018}. The eigenvectors corresponding to the propagating optical fields of Eq.~\eqref{N_matrix_eqn} are given by,
\begin{equation}\label{eigenvector_plus}
    \begin{pmatrix}
    E_x\\
    E_y
    \end{pmatrix}_{+} = |{\bf E}(\omega)| A_+ \begin{pmatrix}
    \chi_{xy} + \chi^{(2)}_{xyy}E^d_y\\
    {\Theta_{+} - \chi^{(2)}_{xxy} E^d_y}
    \end{pmatrix}~, 
    \end{equation}
and, 
\begin{equation} \label{eigenvector_minus}
    \begin{pmatrix}
    E_x\\
    E_y
    \end{pmatrix}_{-} = |{\bf E}(\omega)| A_- \begin{pmatrix}
    {\Theta_{-} - \chi^{(2)}_{yyy} E^d_y} - \eta^2 N^2\\
    \chi_{yx} + \chi^{(2)}_{yyy}E^d_y
    \end{pmatrix}~.
    \end{equation}
Here, $|{\bf E}(\omega)|$ is the strength of the incident field, and $A_\pm = A_\pm (\omega)$, are multiplicative factors normalizing the eigenvectors. These eigenvectors are tunable and can be electrically controlled by varying $E^d_y$. 

An interesting consequence of the unequal refractive indices is the induced polarization rotation, $\Delta \Tilde{\phi},$ for a linearly polarized light incident on the medium, and traverses through it. It can be quantified in terms of the finite birefringence given by $\Delta \tilde{n} = {\rm Re}[\tilde{N}_+ - \tilde{N}_-]$. The linearly polarised light splits into two circularly polarised (CP) or elliptically polarised (EP) lights of opposite chirality upon entering the medium. 

Consequently, upon exiting the medium, these superimpose into a linearly polarised light, with a rotated plane of polarisation w.r.t. the incident light. This frequency-dependent polarization rotation is given by   \cite{Li2022,Cui2014},
\begin{equation}\label{phase rotation}
    \Delta \tilde{\phi} (\omega) = \Delta \tilde{n}(\omega) \frac{\omega L}{c}~. 
\end{equation}
Here, $c$ is the speed of light in free space, and $L$ is the length of the light path through the sample. This phenomenon is similar to the polarization rotation caused by the optical activity of chiral molecules \cite{landau2013}. However, in this case, the optical activity is controlled by a dc electric field. 

In addition to the change in the refractive index, a static electric field can also induce a change in the absorption coefficient. This is instrumental for modulating the intensity of the propagating optical beams. Additionally, the two different absorption coefficients give rise to a frequency-dependent ellipticity of the propagating light and are also related to the dichroism. Analogous to the polarization rotation, the ellipticity ($\tan \varepsilon$) is given by \cite{visnovsky2018},
\begin{equation}
    \tan \varepsilon = - \tanh \left[ \frac{\omega L}{2c} \Delta \tilde{\kappa} \right]~,
\end{equation}
where  we have defined $\Delta \tilde{\kappa} = (\tilde{\kappa}_+ - \tilde{\kappa}_{-}) = \text{Im} [\tilde{N}_{+} - \tilde{N}_{-}]$. 

In this section, we have developed a general theory on birefringence arising from the impact of a static dc field captured via the nonlinear conductivity. In the next section, we focus on the linear EO effect.

 \section{Linear Electro-optic effect}\label{section2.5}

 The static electric-field dependent refractive indices obtained in Eq.~\eqref{Ntilde_solved exprn}, is a nonlinear function of $E_y^d$. Here, we focus on Pockels effect, which captures the linear $E_y^d$ contribution to the polarization rotation.  For this, we expand the altered complex RI in Eq.~\eqref{Ntilde_solved exprn} in powers of $E_y^d$, to obtain 
\bea\label{zeta_expn} 
\zeta=\zeta^{(0)}+\zeta^{(1)}E^d_y~.
\eea 
Here, $\zeta^{(0)}=\eta^2N^2$ and $\zeta^{(1)}=(\chi_{yyy} + \chi_{xxy})$. Similarly, we have, 
\bea \label{xi_expn}
\xi=\xi^{(0)}+\xi^{(1)}E_y^d+\xi^{(2)}(E_y^d)^2~,
\eea 
where $\xi^{(0)}=-\chi_{xy}\chi_{yx}$, $\xi^{(1)}=(\eta^2 N^2\chi_{xxy} - \chi_{xy} \chi_{yxy}-\chi_{yx}\chi_{xyy})$, and $\xi^{(2)}=(\chi_{xxy}\chi_{yyy}+\chi_{xyy}\chi_{yxy})$. These expansions allow us to express $\tilde{N}_{\pm}$ up to the first order of $E_y^d$ as,
\begin{equation} 
    \tilde{N}_{\pm} = \frac{\sqrt{2N^2  +\zeta^{(0)}\pm \rho}}{\sqrt{2}} + \frac{\zeta^{(1)}\left(\rho \pm \zeta^{(0)}\right) \mp 2\xi^{(1)}}{2\sqrt{2} \rho \sqrt{2N^2+\zeta^{(0)} \pm \rho}}E_y^d~,
\end{equation}
with, $\rho=\sqrt{[\zeta^{(0)}]^2-4\xi^{(0)}}$. The difference in complex RI arising from the Pockels effect is given by, 
\begin{widetext}
\begin{eqnarray}
    \Delta \tilde{N}=\tilde{N}_{+}-\tilde{N}_{-} &=&\bigg[\frac{\zeta^{(1)}\left(\rho+\zeta^{(0)}\right)-2\xi^{(1)}}{2\sqrt{2} \rho \sqrt{2N^2+\zeta^{(0)}+\rho}} - \frac{\zeta^{(1)}\left(\rho-\zeta^{(0)}\right)+2\xi^{(1)}}{2\sqrt{2} \rho \sqrt{2N^2+\zeta^{(0)}-\rho}} \bigg]E^d_y ~.
\end{eqnarray}
\end{widetext}
The real and imaginary parts of the above expression capture the birefringence ($\Delta\tilde{n}$) and dichroism ($\Delta\tilde{\kappa}$) induced by the applied static electric field to first order. To evaluate these effects, we compute the second-order susceptibilities $\chi^{(2)}_{abc}(\omega;\omega,0)$. The following section will detail this calculation using the density matrix-based quantum kinetic theory.

\section{second order Optical responses} \label{section3}
In this section, we calculate the components of the second-order susceptibility $\chi_{abc}^{(2)}(\omega;\omega,0)$ using the 
density-matrix based quantum kinetic theory formalism. We will include the impact of a finite Fermi surface and disorder in the second-order responses, which are usually neglected~\cite{sipe2000, Li2022}.  We use the quantum Liouville equation (QLE) to calculate the non-equilibrium density matrix. The time evolution of the density matrix is given by, 
\be \label{QLE} 
\dfrac{\partial \rho(\bm k, t)}{\partial t} + \frac{i}{\hbar}[{\cal H}, \rho(\bm k, t)] = 0 ~.
\ee 
Here, $\rho = \rho(\bm k, t)$ is the non-equilibrium density matrix, and ${\cal H} = \hat{H}_0 + \hat{H}_{\rm E}$, with $\hat{H}_0$ describing the unperturbed Bloch Hamiltonian and $\hat{H}_{\rm E}$ is the perturbation Hamiltonian capturing the effect of the electric field. Using the dipole approximation in the length gauge \cite{Aversa1995, Taghizadeh2017} framework, the light-matter interaction term can be expressed as,  
\be
\hat{H}_{\rm E} = e{\bm E}(t) \cdot \hat{\bm r}~.
\ee
Here, $\hat{\bm r}$ is the position operator, and ${\bm E}(t)$ is the time-dependent electric field. We perturbatively solve for the density matrix up to second order in the electric field strength.  To solve Eq.~\eqref{QLE}, we use the adiabatic switching on of the perturbing fields, ${\bm E}(t) = {\bm E}e^{-i(\omega+i/\tau) t}$. This approach, combined with the interaction picture, has been extensively used in literature for the perturbative solution of the density matrix~\cite{boyd2003, Das2023}. Incorporating such an adiabatic switching on approximation gives rise to a relaxation term in the kinetic equation. For the $N$th order density matrix ($\rho^{(N)} \propto E^N$), Eq.~\eqref{QLE} yields \cite{Bhalla2022, mandal2024, Varshney2023},
\be \label{recursive_dm}
\dfrac{\partial \rho^{(N)}}{\partial t} + \dfrac{i}{\hbar}[\hat{H}_0, \rho^{(N)}] + \dfrac{\rho^{(N)}}{\tau/N} 
= \dfrac{-ie{\bm E}}{\hbar}\cdot[\bm{\hat r}, \rho^{(N-1)}] ~.
\ee
In the above expression, we consider the relaxation time $\tau$ to be a constant. It accounts for the scattering of electrons with static and dynamic (phonons) impurities. We use this equation to obtain the density matrix of various orders in the electric field, as shown in Appendix~\ref{App_A}. Equipped with the expressions of the density matrices, we calculate the NL electric current  using the expression, 
\be \label{curr_def}
{\bm j}= -e {\rm Tr}[{\rho}\hat{\bm v}]~.
\ee
Here, the velocity operator matrix, expressed in the eigenbasis of $\hat H_0$, is given by  $v_{pm}^a=v_m^a \delta_{pm} + (1-\delta_{pm}) i\omega_{pm} \mathcal{R}_{pm}^a\mathcal{R}_{pm}^a$, where $\mathcal{R}_{pm}^a = \langle u_{\bm{k},p}| \partial_{k_a} u_{\bm{k},m} \rangle$ with $|u_{\bm{k},m} \rangle$ being the $m$th band Bloch electron's wavefunction, is the interband Berry connection. From the current density, the second-order conductivities can be extracted via the relation,
\be \label{curr_3rd_def}
j_a^{(2)}(\omega) = \sigma_{abc}^{(2)}(\omega;\omega,0) E_b(\omega) E_c~.
\ee
We obtain the second-order susceptibility from the second-order conductivity using the relation,
\be 
\chi^{(2)}_{abc}(\omega; \omega,0) = \frac{i}{\omega \epsilon_0}\sigma^{(2)}_{abc} (\omega; \omega,0)~. 
\ee
\begin{figure*}[t!] 
    \includegraphics[width=0.8\linewidth]{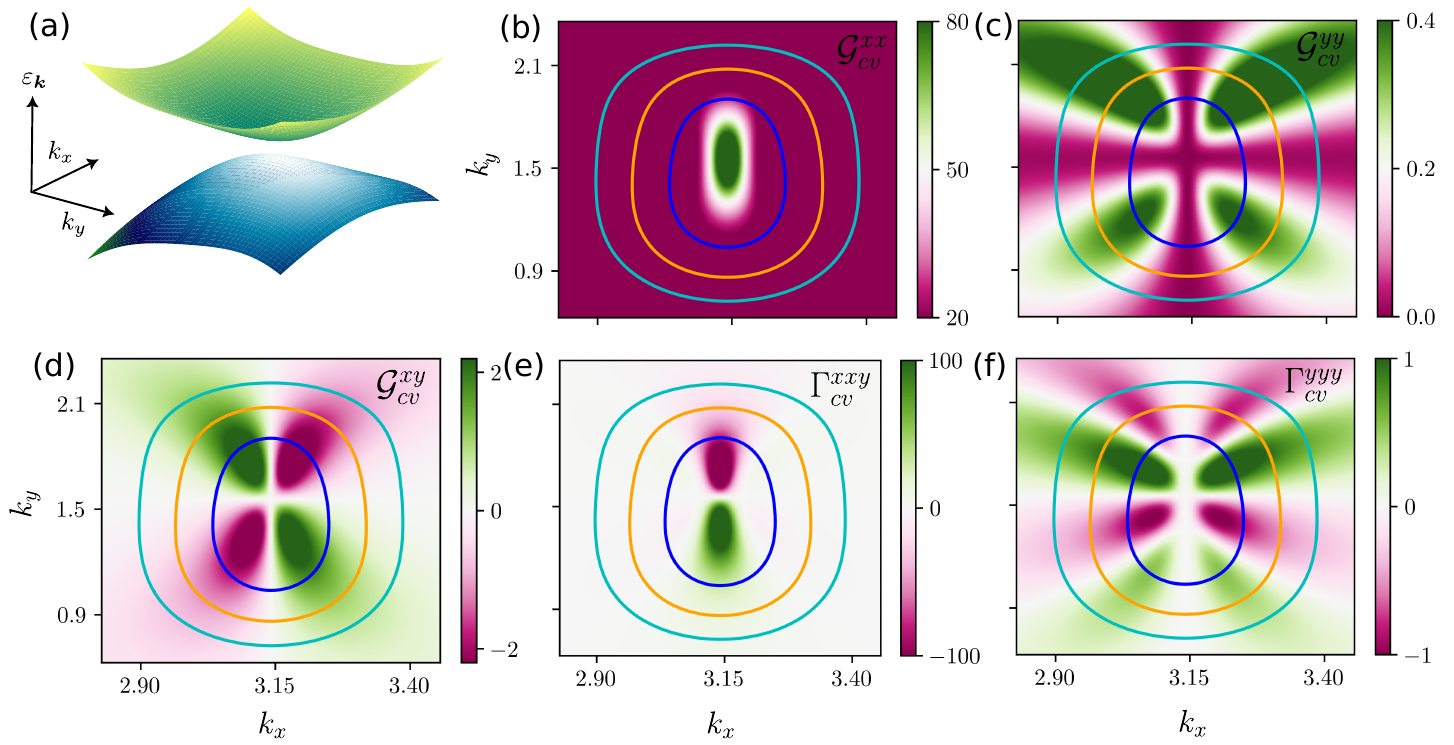}
    \caption{(a) Band dispersion for two dimensional $\mathcal{PT}$-symmetric CuMnAs. We set the hopping parameters as t = 0.08 eV and $\tilde{t} = 1$ eV. The other parameters are $\alpha_R= 0.8$, $\alpha_D= 0$, and ${\bm f}_{AF}=(0.85,0,0)$ eV. The momentum space distribution of the quantum metric components: (b) $\mathcal{G}^{xx}_{cv}$, (c) $\mathcal{G}^{yy}_{cv}$, and (d) $\mathcal{G}^{xy}_{cv}$. The momentum space distribution of the metric connection (e) $\Gamma^{xxy}_{cv}$ and (f) $\Gamma^{yyy}_{cv}$. We have set the temperature to be $T = 12$K, and the scattering time $\tau = 10^{-14}$ s for all our calculations.  
\label{density_plot}}
\end{figure*}
Following Ref.~\cite{Bhalla2022}, we express the second-order susceptibility into six distinct forms based on their origin. We denote these contributions as the shift ($\chi_{abc}^{\rm Sh}$), injection ($\chi_{abc}^{\rm In}$), Drude ($\chi_{abc}^{\rm D}$), anomalous ($\chi_{abc}^{\rm An}$), higher-order pole ($\chi_{abc}^{\rm HOP}$), and double resonant ($\chi_{abc}^{\rm DR}$) susceptibility, respectively. The shift and the injection currents arise from electron position and electron velocity transition (along current direction) during an optical interband excitation  \cite{Ahn2020, Han2024}. These are both Fermi sea effects. The NL Drude contribution arises from the intraband transition and is a Fermi surface contribution \cite{ bhalla2022_prb}.  The anomalous and the double resonant currents originate from the Fermi surface effects. The higher-order pole contribution occurs due to the velocity injection along the applied field direction, and it is a Fermi sea phenomenon. The NL susceptibilities in Eq. \eqref{Ntilde_solved exprn} is an aggregate of all these contributions. In addition to the contributions from the shift and injection currents \cite{Sonowal2019}, we highlight additional contributions to the NL conductivity  \cite{Bhalla2022}. The pairwise field-symmetrized NL susceptibilities are given by,
\begin{widetext}
\begin{subequations}
\begin{eqnarray}\label{susceptibilities}
        \chi^{{\text{Sh}} }_{abc} &=& -\frac{ie^3}{2\omega \epsilon_o \hbar^2}\sum_{m,p} \int_{\rm BZ} [d\bm{k}]~ \omega_{mp} \tilde{g}^\omega_{mp} \left[g^\omega_{mp}\left(\Gamma^{abc}_{mp} - i \tilde{\Gamma}^{abc}_{mp}\right) + g^0_{mp}\left(\Gamma^{acb}_{mp} - i \tilde{\Gamma}^{acb}_{mp}\right)\right]F_{mp}~,\\
        \chi^{{\text{Inj}} }_{abc} &=&\frac{ie^3}{2 \epsilon_0\omega\hbar^2} \sum_{m,p} \int_{\rm BZ} [d\bm{k}]~ \partial_a(\omega_{mp})g^\omega_0\left[ g^\omega_{mp}\left(\mathcal{G}^{bc}_{mp} -i \frac{\Omega^{bc}_{mp}}{2} \right) + g^0_{mp}\left(\mathcal{G}^{cb}_{mp} -i \frac{\Omega^{cb}_{mp}}{2} \right)\right]F_{mp}~,\\
        \chi^{{\text{D}} }_{abc} &=& -\frac{ie^3}{2 \epsilon_0\omega\hbar^2} \sum_{m} \int_{\rm BZ} [d\bm{k}]~\left[\frac{1}{\hbar} \partial_a \epsilon_{\textbf{k},m} (\tilde{g}^\omega_0 {g}^\omega_0 + \tilde{g}^0_0 {g}^\omega_0) \partial_b \partial_c f_m\right]~, \\    \chi^{{\text{An}} }_{abc} &=& -\frac{ie^3}{2 \epsilon_0\omega\hbar^2} \sum_{m,p}\int_{\rm BZ} [d\bm{k}]~\omega_{mp}g^\omega_{mp}\left[g^{\omega}_0\left(\mathcal{G}^{ab}_{mp} -i \frac{\Omega^{ab}_{mp}}{2} \right)\partial_cF_{mp}+ g^0_0\left(\mathcal{G}^{ac}_{mp} -i \frac{\Omega^{ac}_{mp}}{2} \right)\partial_bF_{mp}\right]~,\\
        \chi^{{\text{HOP}} }_{abc} &=& -\frac{ie^3}{2 \epsilon_0 \omega \hbar^2}\sum_{m,p}\int_{\rm BZ} [d\bm{k}]~ \omega_{mp}g^\omega_{mp}\left[\left(\mathcal{G}^{ac}_{mp} -i \frac{\Omega^{ac}_{mp}}{2} \right)\partial_b g^{\omega}_{mp} +\left(\mathcal{G}^{ab}_{mp} -i \frac{\Omega^{ab}_{mp}}{2} \right)\partial_c g^0_{mp}\right]F_{mp}~,\\
        \chi^{{\text{DR}} }_{abc} &=& -\frac{ie^3}{2 \epsilon_0 \omega \hbar^2}\sum_{m,p} \int_{\rm BZ} [d\bm{k}]~\omega_{mp}g^\omega_{mp}\left[ g^\omega_{mp}\left(\mathcal{G}^{ac}_{mp} -i \frac{\Omega^{ac}_{mp}}{2} \right)\partial_bF_{mp} + g^0_{mp}\left(\mathcal{G}^{ab}_{mp} -i \frac{\Omega^{ab}_{mp}}{2} \right)\partial_cF_{mp} \right]~.
    \end{eqnarray}
\end{subequations}
\end{widetext}
In the above expressions, $[d\textbf{k}] = {d^dk}/{(2\pi)^d}$ in $d$ spatial dimensions. The NL susceptibilities are primarily determined by four band geometric quantities: the Berry curvature ($\Omega$), the quantum metric ($\mathcal{G}$), the metric connection ($\Gamma$), and the symplectic connection ($\tilde{\Gamma}$) \cite{Xiao2010, Kozii2021, RADU1996, mandal2024}. The former two are the real and imaginary parts of the QGT ($ \mathcal{Q}=\mathcal{G}-i\Omega/2$), while the latter two are the real and imaginary parts of the QGC ($ \mathcal{C}=\Gamma-i\tilde{\Gamma}$), respectively. The functions, $g^{\omega}_{mp} = [1/\tau - i (\omega - \omega_{mp})]^{-1}$ and $\tilde{g}^{\omega}_{mp} = [2/\tau - i (\omega - \omega_{mp})]^{-1}$ with $\hbar\omega_{mp} = (\epsilon_{m,\textbf{k}} - \epsilon_{p, \textbf{k}}) $ are related to the joint density of states broadened by disorder. $F_{mp} = f^{(0)}_m - f^{(0)}_p$ is the difference in the occupation of the $m$ and $p$ bands in equilibrium. The Fermi-Dirac distribution function gives the normalized occupation density in the $m$th band, $f^{(0)}_{m} = [1 + e^{\beta( \epsilon_{m,\textbf{k}} - \mu)}]^{-1}$ with $\beta (= 1/k_BT)$. Here, $T$, $\mu$, and $k_B$ are the thermodynamic temperature, chemical potential, and the Boltzmann constant, respectively. 

These NL responses at frequency $\omega$, combined with Eq.~\eqref{Ntilde_solved exprn} complete the theoretical framework of electric field-induced modulation of the RI of a material or the electro-optic effect. The results obtained so far are general and not constrained by any symmetry restrictions. However, different band geometric quantities can either dominate or vanish in systems with specific symmetries, making the induced EO effect an indirect probe of the band geometry of the system. An interesting system is where both $\mathcal{P}$ and $\mathcal{T}$ symmetries are broken, while their combined action, $\mathcal{PT}$, remains intact \cite{Xia2021, Klauck2019, Kremer2019, Tiwari2020}. In ${\cal P}{\cal T}$ symmetric systems, the Berry curvature and the symplectic connection vanish over the entire Brillouin zone \cite{Bhalla2022}. As a consequence, the electro-optic effect is induced entirely by the quantum metric and the metric connection. The linear order anomalous Hall response also vanishes as the Berry curvature is zero over the entire Brillouin zone in ${\cal P}{\cal T}$ symmetric systems. 

In the next section, we apply our theory to a $\mathcal{PT}$-symmetric model. This will allow us to illustrate the contributions of the quantum metric and the metric connection.

\section{Linear Electro-Optic Effect in $\mathcal{PT}$ - symmetric systems}\label{section4} 

\begin{figure}[t!]
    \centering
    \includegraphics[width=\linewidth]{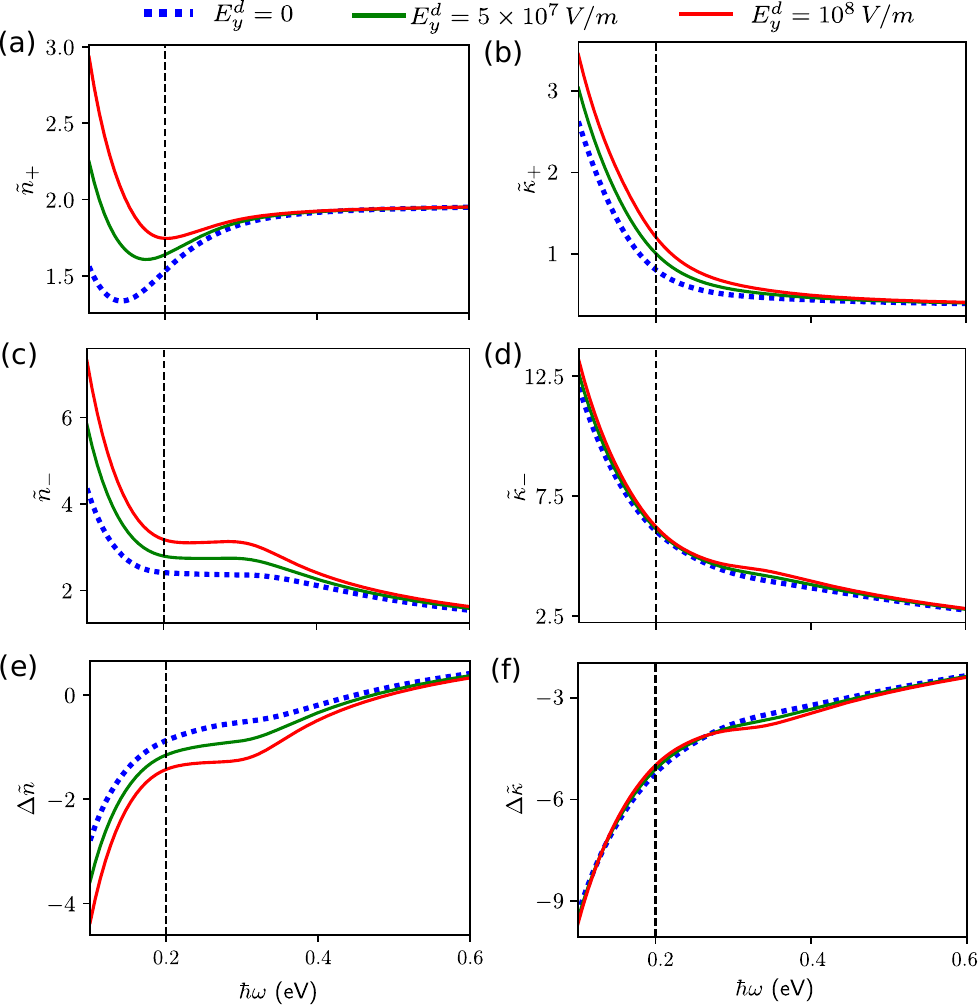}
   \caption{The frequency dependence of the refractivities (a) $\tilde{n}_{+}$ and c) $\tilde{n}_{-}$, respectively. (b) and (d) depict the frequency dependence of the absorption coefficients, $\tilde{\kappa}_{+}$ and $\tilde{\kappa}_{-}$, respectively. The variation of (e) $\Delta\tilde{n}$ and (f) $\Delta\tilde{\kappa}$ with frequency. The response for different values of the dc electric field, $\bm{E}^d = E^d_y \hat{y}$, in (a)-(f) are indicated by the blue dots ($E^d_y=0$), green line ($E^d_y=5 \times 10^7$ V/m), and red the line ($E^d_y = 10^8$ V/m) respectively. The dotted vertical line represents $\hbar \omega=\mu$, where $\mu$ is the chemical potential. For all our calculations, we have set the temperature to be $T = 12$ K and the scattering time $\tau$ to be $10^{-14}$ s.}\label{RI_plot}
\end{figure}

In this section, we study the linear EO effect in CuMnAs, a  $\mathcal{PT}$ symmetric antiferromagnet with broken $\mathcal{P}$ and $\mathcal{T}$ symmetry \cite{Tang2016, Cao2023, Rui2019, Wadley2016}. CuMnAs hosts oppositely magnetized sublattices (or Mn atoms - designated as $A$ and $B$), which break the $\mathcal P$ and $\mathcal T$ symmetry. However, the simultaneous interchange of the $A$ and $B$ sublattices with a magnetization flip keeps the system unchanged, reflecting its ${\mathcal P}{\mathcal T}$ symmetry. The opposite magnetization also leads to sublattice-dependent anti-symmetric spin-orbit coupling. The effective tight-binding Hamiltonian of this system \cite{Watanabe2021, Das2023} is given by

\begin{equation} \label{cumnas_ham}
    \mathcal{H} (\textbf{k}) =\begin{pmatrix}
    \epsilon_0(\textbf{k}) + \textbf{f}_A(\textbf{k})\cdot {\bm \sigma} && V_{AB}(\textbf{k}) \\
    V_{AB}(\textbf{k}) && \epsilon_0(\textbf{k}) + \textbf{f}_B(\textbf{k})\cdot{\bm \sigma}
    \end{pmatrix}~.
\end{equation}
Here, we have $ \epsilon_0(\textbf{k}) = -t(\cos k_x + \cos k_y)$, $ V_{AB} = -2\tilde{t} \cos \frac{k_x}{2} \cos \frac{k_y}{2}$, and ${\bm \sigma}$ represents the Pauli spin matrices. The hopping between orbitals of the same sublattice is quantified by $t$, while $\tilde{t}$ denotes the hopping between orbitals on different sublattices. The sublattice-dependent spin-orbit coupling and the magnetization field are included in ${\bm f}_A({\bm k})=-{\bm f}_B({\bm k})$ and we have,
\begin{align}
    \bm{f}_A(\bm{k}) 
    = \begin{pmatrix}
    f^x_{AF} - \alpha_r \sin(k_y) + \alpha_d \sin (k_y) \\
    f^y_{AF} + \alpha_r \sin(k_x) + \alpha_d \sin(k_x) \\
    f^z_{AF} \\
    \end{pmatrix}~.
\end{align}
The electronic band dispersion of the Hamiltonian in Eq.~\eqref{cumnas_ham} are given by  $\epsilon (\textbf{k}) = \epsilon_0 \pm \sqrt{V^2_{ab} + f^2_{Ax} + f^2_{Ay} +f^2_{Az}}$, and shown in Fig.~\ref{density_plot}(a). Here, $+(-)$ denotes the conduction (valence) band. We present the momentum space distribution of different components of the quantum metric and the metric connection in Fig.~\ref{density_plot}(b)-(f). These components are pivotal for inducing polarization rotation within $\mathcal{PT}$-symmetric systems. Fig.~\ref{susceptibility_plot} in Appendix~\ref{App_A}  shows the real and imaginary parts of $\chi_{xxy}$ and $\chi_{yyy}$ for CuMnAs. As the Berry curvature and symplectic connection vanish for a $\mathcal{PT}$ symmetric system,  the shift current is induced by the metric connection, while the other contributions arise from the quantum metric. For CuMnAs, all the contributions of $\chi_{xyy}$ and $\chi_{yxy}$ components are zero.

In Fig.~\ref{RI_plot}(a) and \ref{RI_plot}(b), we show the variation of $\tilde{n}_+$ and $\tilde{\kappa}_+$ with optical frequency.  Figure~\ref{RI_plot}(c) and  \ref{RI_plot}(d) show the variation of $\tilde{n}_-$ and $\tilde{\kappa}_-$ with frequency.  Since the  NL susceptibilities are approximately seven orders smaller than the linear susceptibility, the effect of the dc electric field becomes significant around $E^d_y = 10^{7}$ V/m or so. Experimentally, external field ($E^d_y = 10^6$ V/m) induced linear EO effect (Pockels effect) has been observed in ferroelectric materials like C$_8$H$_8$N$_2$ and LiBNO$_3$ with birefringence being of the order of $\sim 10^{-5}$ and $\sim 10^{-4}$ respectively \cite{Uemura2020}. We present the altered birefringence, $\Delta \tilde{n} = \Delta \tilde{n}_{+} - \Delta \tilde{n}_{-} ,$ as a function of frequency in Fig.~\ref{RI_plot}(e). In Fig.~\ref{RI_plot}(f), we present the dichroism, $\Delta \tilde{\kappa},$ as a function of frequency. We note that in Fig.~\ref{susceptibility_plot}, all the non-monotonic variation of the optical susceptibilities occur for frequencies $\hbar \omega \ll 2 \mu$. Beyond the frequencies $\hbar \omega \ll 2 \mu$, the nonlinear susceptibilities vanish as $1/\omega$. A similar trend can also be seen in all panels of Fig.~\ref{RI_plot}.

There are two distinct differences between the electro-optic effect in conventional ferroelectric materials and ${\cal P}{\cal T}$ symmetric systems. In contrast to ferroelectric materials, the EO effect in CuMnAs is induced by the band geometric quantities, such as the quantum metric and the metric connection. Additionally, CuMnAs supports the EO effect for optical energies below the chemical potential [see the region to the left of the dashed vertical line for $\mu = 0.2$ eV in Fig.~\ref{RI_plot}]. This is a distinct signature of Fermi surface effects playing a significant role in the total optical susceptibility in metallic systems. 

\section{Conclusion}\label{endgame}
To summarize, our study highlights the intricate phenomena of polarization rotation induced by linear electro-optic effect-induced alteration in their refractive index. 
Using quantum kinetic theory and density matrix formalism, we have identified six distinct contributions to the linear EO effect: shift, injection, Drude, anomalous, higher-order pole, and double resonant currents. Our work emphasizes the significant roles of the quantum metric and metric connection in $\mathcal{PT}$-symmetric systems, where Berry curvature and symplectic connection are nullified. We derived expressions for the change in refractive index (RI) and calculated the resulting polarization rotation of emergent light in the presence of a transverse static electric field.

From an application perspective, the electro-optic effect is the cornerstone for electro-optic modulation, which is crucial for precise control of optical parameters. Electro-optic modulators are indispensable optical components for modulating light's intensity (intensity modulators) and phase or polarization (phase modulators) \cite{Jeffrey2004, Forouzmand2019, Zhang2019, Chen2014, Li2020, Wang2018}. In addition to serving as a potential probe for band geometric quantities, efficient EO modulators drive innovation across the diverse fields of optical communications \cite{Wooten2000, Chelmus2022}, computing \cite{Portner2021}, microwave \cite{Marpaung2019} and quantum photonics \cite{Sun2015, David2018, Karpinski2017}, sensing \cite{Benea2022, Prasad2017}, and telecommunications \cite{dalton2019, Kumar2015, Krasavin2012}.

\section{Acknowledgement}
We acknowledge Debasis Dutta, Kamal Das, and Debottam Mandal for the useful discussions. M.M.K. acknowledges the Department of Science and Technology, Government of India, for financial support via Project No. DST/NM/TUE/QM-6/2019(G)-IIT Kanpur.
S. S. thanks the MHRD, India, for funding through the Prime Minister’s Research Fellowship (PMRF).
\appendix
\section{Optical susceptibility calculations}\label{App_A}
In this section, we provide a detailed calculation of the optical conductivity using the density matrix up to the second order in the electric field. Perturbatively treating the density matrix in the quantum Liouville equation, the $N$th order density matrix can be shown to be related to the $(N-1)$th order density matrix via the recursive relation, 
\be \label{QKE_nth_order}
\dfrac{\partial \rho^{(N)}_{mp}}{\partial t} + \dfrac{i}{\hbar}[\hat{H}_0 , \rho^{(N)}]_{mp} + \dfrac{\rho^{(N)}_{mp}}{\tau/N} = \frac{-ie\bm E}{\hbar}\cdot [{\bm r}, \rho^{(N-1)}]_{mp}~.
\ee 
In the steady state, this simplifies to, 
\be
\left(-i\omega_\Sigma + i\omega_{mp} + \dfrac{N}{\tau} \right) \rho^{(N)}_{mp} = \dfrac{e{\bm E}}{\hbar}\cdot [\mathcal{D}_{\bm k}\rho^{(N-1)}]_{mp} ~,
\ee 
which leads us to the recursive equation for $\rho^{(N)}$,
\be \label{recursive_nth_order}
\rho^{(N)}_{mp} = \frac{e}{\hbar} g_{mp}^{\omega_\Sigma} {\bm E} \cdot [\mathcal{D}_{\bm k}\rho^{(N-1)}]_{mp} ~.
\ee 
The function $g_{mp}^{\omega_\Sigma}=[N/\tau - i(\omega_\Sigma - \omega_{mp})]^{-1}$ characterizes the density of states broadened by scattering corresponding to incident frequency $\omega_\Sigma$. 
Here, $\omega_\Sigma$ and $\omega_{mp} = \hbar^{-1}(\epsilon_{m, \bm{k}} - \epsilon_{n, \bm{k}})$ are the sum of the harmonic frequencies incident on the system, and difference of the frequencies corresponding to bands indexed $m$ and $n,$ respectively.  The first-order density matrix is given by, 
\be \label{1st_order_dm}
\rho^{(1)}_{mp} = \frac{e}{\hbar} g_{mp}^{\omega} {\bm E} \cdot \left( \partial_{\bm k}\rho_{mp}^{(0)} - i[\mathcal{R}_{\bm k },\rho^{(0)}]_{mp} \right) ~. 
\ee
Here, $\rho^{(0)}_{mp} = f_m~\delta_{mp}$ represents the equilibrium distribution of electrons, with $f_m = [1+ e^{\beta(\epsilon_{m{\bm k}}-\mu)}]^{-1}$ being the Fermi-Dirac distribution for the $m$th band, and $\beta  = 1/(k_BT)$. With some simplifications, the first-order density matrix can be expressed as, 
\be \label{1st_order_dm_total}
\rho^{(1)}_{mp} = \frac{e}{\hbar} g^\omega_{mp}(\partial_c f_m + i \mathcal{R}^c_{mp}F_{mp})E_c e^{-i\omega t} ~.
\ee 
We note that for $m = p$, the second term of the above equation vanishes, and for $m \neq p$ the first term vanishes. 
Calculating the second-order density matrix, we obtain,  
\begin{widetext}
\be \label{2nd_order_dm}
\rho^{(2)}_{mp} = \frac{e}{\hbar} g_{mp}^{\omega_\Sigma} {\bm E} \cdot \left[ \partial_{\bm k}\rho_{mp}^{(1)} - i[\mathcal{R}_{\bm k },\rho^{(1)}]_{mp} \right] ~
\\
= \frac{e}{\hbar} g_{mp}^{\omega_\Sigma} E_c \left[ \partial_c \rho_{mp}^{(1)} - i\sum_n \left( \mathcal{R}^c_{mn}\rho_{np}^{(1)} - \rho_{mn}^{(1)}\mathcal{R}^c_{np}\right) \right] ~.
\ee
%
\begin{figure}[h!] 
    \includegraphics[width=0.8\linewidth]{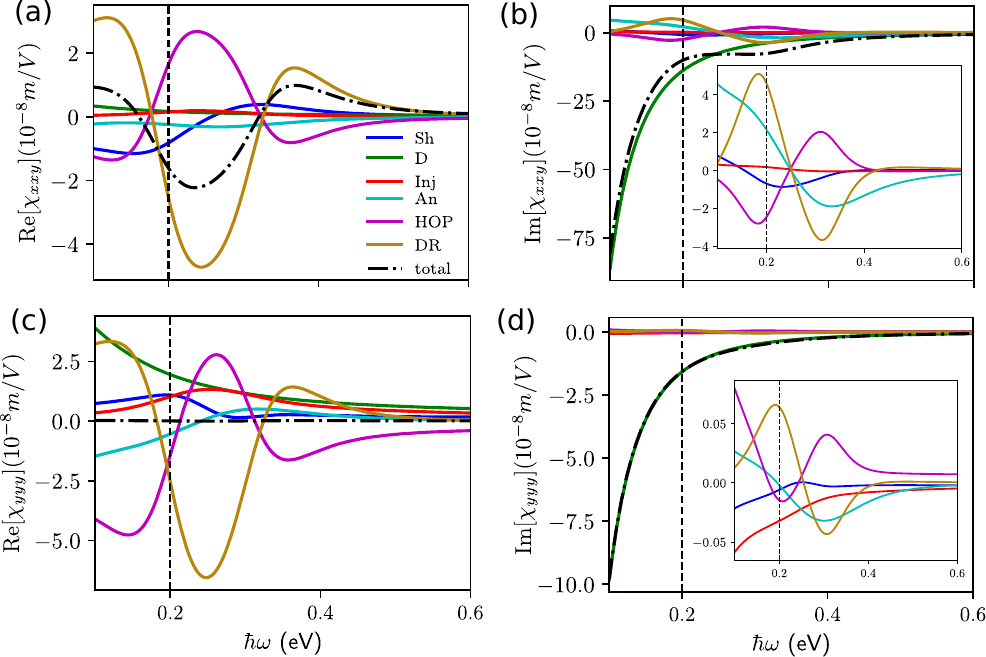}
    \caption{Frequency dependence of various components - shift (Sh), Drude (D), injection (Inj), anomalous (An), higher-order pole (HOP), and double-resonant (DR), of nonlinear susceptibility. Panels (a) and (b) show the real and imaginary parts of the susceptibility $\chi_{xxy}$, respectively. Panels (c) and (d) show the real and imaginary parts of the susceptibility $\chi_{yyy}$, respectively. The susceptibilities without the Drude contribution are shown in the inset of panels (b) and (d). The model parameters are the same as in fig.~\ref{density_plot}. }        
   \label{susceptibility_plot}
\end{figure}

For a two-band model, the second-order density matrix can be separated into four parts: two diagonal and two off-diagonal terms. We define them as, $\rho_{mm}^{(2),I}$, $\rho_{mm}^{(2),II}$, $\rho_{mp}^{(2),I}$, and $\rho_{mp}^{(2),II}$, respectively. These four parts after pairwise field symmetrization can be expressed as, 
\bea 
\rho_{mm}^{(2),I}(\omega;\omega,0)
&=& \dfrac{e^2}{\hbar^2}  (\tilde{g}^\omega_0 {g}^\omega_0 + \tilde{g}^0_0 {g}^\omega_0) \partial_c\partial_d f_m E_c E^d_y~,\label{rho_dd} \\
\rho_{mm}^{(2),II}(\omega;\omega,0)
&=& \dfrac{e^2}{\hbar^2} \tilde{g}^\omega_0 \sum_n[({g}^\omega_{nm}\mathcal{R}^b_{mn} \mathcal{R}^c_{nm}F_{nm} - {g}^\omega_{mn} \mathcal{R}^c_{mn} \mathcal{R}^b_{nm}F_{mn})+({g}^0_{nm}\mathcal{R}^c_{mn} \mathcal{R}^b_{nm}F_{nm} - {g}^0_{mn} \mathcal{R}^b_{mn} \mathcal{R}^c_{nm}F_{mn})]~,\label{rho_od} \nonumber\\
\\
\rho_{mp}^{(2),I}(\omega;\omega,0)
&=& i\dfrac{e^2}{\hbar^2} \tilde{g}^\omega_{mp} \left[\left({g}^\omega_{mp}F_{mp}\mathcal{D}^b_{mp} \mathcal{R}^c_{mp} + \mathcal{R}^c_{mp}\partial_b({g}^\omega_{mp}F_{mp})\right)+\left({g}^0_{mp}F_{mp}\mathcal{D}^c_{mp} \mathcal{R}^b_{mp} + \mathcal{R}^b_{mp}\partial_c({g}^0_{mp}F_{mp})\right)\right]~, \label{rho_do}\\
\rho_{mp}^{(2),II}(\omega;\omega,0)
&=& i\dfrac{e^2}{\hbar^2} \tilde{g}^\omega_{mp} \left[{g}^\omega_{0}\mathcal{R}^b_{mp}\partial_c F_{mp} +{g}^0_{0} \mathcal{R}^c_{mp}\partial_bF_{mp}\right]~.\label{rho_oo}
\eea
In the above equations, we have considered the frequency combination of $(\omega,0)$.
Further, we calculate the current densities using the definition, 
\begin{equation}
    {\bm j}(t) = -e {\rm Tr}[{\rho}\hat{\bm v}]~.
\end{equation}
So, the $N$th order current can be calculated as,
\begin{equation}\label{nth_current}
    j^{(N)}_a (t) = -e \sum_{m,p}v_{pm}^a\rho_{mp}^{(N)}(t)~.
\end{equation}
Here, $v^a_{pm} = v^{0a}_{pm}\delta_{pm} +i \omega_{pm} \mathcal{R}^a_{pm}$ is the general velocity operator with $v^{0a}_{mm} = \hbar^{-1}\partial_a \epsilon_m,$ and $\mathcal{R}^q_{pm}$ is the $q$th component of the interband Berry connection. Using Eq. \eqref{nth_current}, for $N=1$, we can write the linear current as, 
\begin{align}
    j^{(1)}_a (t) ={}& -\frac{e^2}{\hbar} \sum_{m,p} \sum_c g^\omega_{mp}( v^{0a}_{pm}\delta_{pm} +i \omega_{mp} \mathcal{R}^a_{pm})  (\partial_c f_m + i\mathcal{R}^c_{mp}F_{mp})E^c e^{-i\omega t}\nn~,\\
    &= - \frac{e^2}{\hbar} \sum_c\bigg[ \sum_{m}  \frac{1}{\hbar} (\partial_a \epsilon_m)(\partial_c f_m)g^\omega_0 - \sum_{m,p} \omega_{mp} g^\omega_{mp} \mathcal{R}^a_{pm} \mathcal{R}^{c}_{mp}F_{mp} \bigg] E^ce^{-i\omega t}\nn ~,\\
    &= - \frac{e^2}{\hbar} \sum_c \bigg[ \sum_{m}  \frac{1}{\hbar} (\partial_a \epsilon_m)(\partial_c f_m)g^\omega_0 - \sum_{m,p} \omega_{mp} g^\omega_{mp} \mathcal{Q}^{ac}_{mp}F_{mp} \bigg] E^ce^{-i\omega t}~.
\end{align} 
\end{widetext}  
The product of two Berry connections has been expressed in terms of the QGT. It 
can be expressed in terms of the band-resolved Berry curvature and the quantum metric as, 

\begin{equation}\label{qgt}    \mathcal{Q}^{bc}_{mp} = \mathcal{R}^b_{pm}\mathcal{R}^c_{mp} = \mathcal{G}^{bc}_{mp} - (i/2)\Omega^{bc}_{mp}~.
\end{equation}
 We have two contributions to the current density. Under $\mathcal{PT}$ symmetry, as the Berry curvature vanishes, the linear conductivity corresponding to each of the contributions can therefore be given as,
\begin{align}
    \sigma^{\text{intra}}_{ac} (\omega) &= -\frac{e^2}{\hbar}\sum_{m} \int[d\textbf{k}] \frac{1}{\hbar} (\partial_a \epsilon_m)g^\omega_0(\partial_c f^{(0)}_m)~, \\
    \sigma^{\text{inter}}_{ac} (\omega)&=\frac{e^2}{\hbar} \sum_{m,p}\int[d\textbf{k}] \omega_{pm} g^\omega_{mp} \mathcal{G}^{ac}_{mp}F_{mp}~. 
\end{align}
The linear conductivity $\sigma_{ac} (\omega)$ is given as,
\begin{equation}
    \sigma_{ac} (\omega) = \sigma^{\text{intra}}_{ac} (\omega) + \sigma^{\text{inter}}_{ac} (\omega)~. 
\end{equation}
The corresponding susceptibility tensor is given from $\sigma_{ac} (\omega) = -i \omega \epsilon_0 \chi_{ac}(\omega)$, where $\epsilon_0$ is the permittivity of vacuum. 
It can be shown that the intraband conductivity is the infamous Drude conductivity \cite{ashcroft1976} for frequency-dependent electric fields and that the interband contribution we see for 
$\sigma^{\text{inter}}_{ac}(\omega)$, is the linear anomalous Hall conductivity \cite{Nagaosa2010}. 
As the Berry curvature vanishes in a $\mathcal{PT}$-symmetric system, the total linear contribution to the interband conductivity would be from the quantum metric. At $N=2$ in Eq.~\eqref{nth_current}, we get the second-order current. The expressions of the pairwise field-symmetrized second-order susceptibilities are shown in the main text. 

\bibliography{ref1} 
\end{document}